\documentstyle[epsf,preprint,aps]{revtex}

\begin{document}

\baselineskip 0pt
\parindent 0pt
\parskip 5pt


ORAL/POSTER REFERENCE: ICF100796OR

\vskip 72pt

\newfont{\titlefont}{cmr12 scaled 1333}
\begin{center}
{\bf \titlefont FORMAL CONSIDERATIONS ABOUT FRACTURE:}

{\bf \titlefont NUCLEATION AND GROWTH}
\end{center}

\vskip 12pt

\begin{center}
{
James P. Sethna
}

\vskip 12pt

{
{Laboratory of Atomic and Solid State Physics,}

{Clark Hall, Cornell University, Ithaca, NY 14853-2501, USA}
}
\end{center}

\vskip 36pt
{\bf ABSTRACT}
\vskip 12pt

In this paper, we'll answer several abstract, formal questions about the
nature of crack growth and nucleation. Bringing a field theory point
of view to fracture illuminates things in what I hope will be an entertaining
way.

Formally, what is the crack nucleation rate? Fracture is an instability of
elastic theory under tension. For thermally nucleated cracks, there is an
analogy with supercooled liquids. Here the crack nucleation rate can be
thought of as an imaginary part of the free energy - giving the decay
rate of the metastable, stretched material. As an amusing consequence,
we can formally calculate the asymptotic form for the high-order nonlinear
elastic coefficients, and explicitly show that elastic theory has zero
radius of convergence.

Formally, can we derive the the crack growth laws from symmetry? We can
describe mixed-mode three-dimensional fracture as a moving curve
in space, decorated with a description of the local crack plane,
and driven by the stress intensity factors along the crack. Imposing
the symmetries gives us the form of the crack growth laws in two
dimensions, explaining the (well-known) fact that cracks under shear
(mode II) abruptly turn until the stress becomes purely extensional
(mode I). The form of the crack growth law in three dimensions will be derived,
and linear stability analysis for mixed-mode fracture will be briefly 
summarized, with connection to factory-roof morphologies.

\vskip 24pt

{\bf KEYWORDS}
\vskip 12pt

fracture, crack, nucleation, growth, thermal, evolution, factory roof

\vskip 24pt

{\bf INTRODUCTION}

\vskip 12pt

In this paper, I will summarize formal work done in collaboration with Jennifer
Hodgdon\cite{Hodgdon} and Alex Buchel\cite{BuchelPRL,BuchelPRE} on nucleation
and growth of cracks in brittle, isotropic, homogeneous materials. This work
is formal in the sense that bring to the problem the perspectives and
approaches of theoretical physics: we'll be asking fundamental questions
that most of the participants in this conference wouldn't think to ask.
Sometimes answering very basic questions can be illuminating; in any case
I hope it will be entertaining. 

\vskip 24pt

{\bf CRACK NUCLEATION}

\vskip 12pt

What is the crack nucleation rate? In a practical context, one may ask what
the failure rate is of a component in a machine: the number of breakages per
unit time. This definition depends on the size and shape of the object and
the external loading. More generally, for a given material under a given load,
there will be a rate of crack formation per unit volume per unit time.

In general, the physics underlying this rate is complicated. Cracks 
always nucleate at inhomogeneities in the material (grain boundaries, surface
heterogeneities, pre-existing microcracks, dislocation tangles). These
heterogeneities usually evolve in time (fatigue, work hardening,
electromigration, void formation). This makes the failure rate history
dependent, and too complicated a problem to start with. So, we'll focus
on brittle fracture, where we perhaps can assume that the material far from
the crack tip does not irreversibly transform.

With this approximation, we have two cases. (1) The important case is when
the failure is due to the inhomogeneities: distributions of
microcracks\cite{Curtin}, distributions of bond 
strengths\cite{BondStrengthRefs}, or other randomness\cite{Nattermann}.
In all of these cases, the rate is number of failures for a given increment of
strain, not per unit time. Either the stress is large enough to trigger failure
or not: a given sample, if it survives the initial loading, will survive
forever. This leads to some elegant theories and simulations, but isn't what
I'm focusing on today. (2) The case we study is when the material is 
homogeneous, and the failure is triggered by a thermal fluctuation. This 
basically never happens, but it's an interesting theoretical problem.

So, fundamentally, what is the fracture nucleation rate for our brittle,
homogeneous material? This is actually a deep theoretical physics question.
The basic problem is defining what is meant at finite temperature by a
material under tension. The lowest energy state (free energy, to be pedantic)
for a large cube of brittle material under external strain can easily be seen
to be broken in two. The elastic energy released is proportional to the
volume; the surface energy cost is proportional to the cross-sectional area;
for fixed external strain there will be a length at which breaking wins.
One should note that this breaking strain in real materials isn't very large:
a cubic meter of iron will prefer to break in two when stretched by 10$\mu$m
(140 $\mu$m if one includes the energy lost due to plastic deformation during
fracture).

Crack nucleation is precisely the mechanism by which a material goes from the
metastable stretched state to the state which is broken in two. There is
a close analogy to the nucleation of crystals in supercooled liquids.
In supercooled liquids\cite{Langer} the liquid is a metastable state with
a decay rate given by the thermal probability of generating a critical nucleus. 
For fracture, the stretched state is a metastable state with a decay rate
given by the thermal probability of generating a crack of the size given by
the Griffiths threshold.

(The analogy even extends to the practicality of the thermal nucleation 
mechanisms. Homogeneous nucleation of crystals in supercooled liquids 
almost never happens: the crystal almost always nucleates on a dust particle,
or on the surface...)

Let's do this calculation explicitly, for a two-dimensional straight crack
of length $\ell$, in a medium under external hydrostatic tension (a pressure
$P<0$). One can solve the elastic equations to find the energy of this crack. 
Let $\alpha$ be the surface energy (energy per unit length, in two dimensions,
of the crack surface), $Y$ be the Youngs modulus, and $\sigma$ be Poisson's
ratio. If we define a critical crack length $\ell_c$ (the 
Griffiths\cite{Griffiths} threshold) by
\begin{equation}
\ell_c={{4 Y \alpha} / {\pi P^2 (1- \sigma^2)}}
\end{equation}
then the energy of the crack is
\begin{equation}
E=2 \alpha \ell -\alpha {\ell^2 / \ell_c}.
\end{equation}
It follows that cracks with $\ell>\ell_c$ will grow to lower their energy,
and that cracks with $\ell<\ell_c$ will heal (at least in our model). The
energy at the top of the barrier is $E(\ell_c) = \alpha \ell_c = 
{4 Y \alpha^2} / (\pi P^2 (1- \sigma^2))$, and the probability for 
this critical crack to form by a thermal fluctuation is suppressed by the
Boltzmann factor $\exp(-E(\ell_c)/k_B T)$.  The fracture nucleation rate
$\Gamma_{failure}$ per unit volume per unit time will be given by some
prefactor times this Boltzmann factor:
\begin{equation}
\Gamma_{failure} = \Gamma_0 \exp(-E(\ell_c)/k_B T).
\end{equation}
(Calculating this prefactor $\Gamma_0$ is quite a subtle
problem\cite{BuchelPRE}.)

This calculation is precisely analogous to the corresponding one for 
a critical nucleus of solid forming in a supercooled liquid\cite{Langer}.
Indeed, the crack as it nucleates will be filled by a gas of vapor atoms
(think of fracture in a piece of dry ice, where the vapor pressure isn't
vanishingly small). The spontaneous fracture rate is precisely the nucleation
rate of a gas forming in a super-stretched crystal.

Remember, though, that we claimed from a fundamental point of view the
difficulty is defining the stretched material. A stretched piece of material
is fundamentally a transient: how do we use equilibrium statistical mechanics
to study what is only a metastable equilibrium? The precise definition of
the stretched, metastable state is as an analytical continuation from the
compressed state\cite{Langer,BuchelPRL}. In particular, suppose we study
the energy
$E(P)$ of the material under positive hydrostatic pressure $P$, where the
material is well defined. If we include the effects of thermal fluctuations
producing small cracks, and then analytically continue $E(P)$ to negative
pressure, one can show that it becomes complex! The decay rate (3) is
proportional to the imaginary part of the energy: 
\begin{equation}
{\rm Im} E(P) = C \Gamma_{failure}(P)
= E_0 \exp\left(-{{4 Y \alpha^2} / {\pi P^2 (1- \sigma^2) k_B T}}\right).
\label{eq:ImE}
\end{equation}
Metastable states in statistical mechanics are defined by an analytic
continuation of the (free) energy. The nucleation rate is defined as a
particular dynamical constant prefactor $\Gamma_0/E_0$ times the imaginary
part of this complex energy.

What can we derive from this formal imaginary part to the energy? The real
part of the energy is related to the bulk modulus? Let's
define the pressure-dependent bulk modulus $K(P)$ by
\begin{equation}
{1 \over K(P)} = -{1 \over V} \left({\partial V \over \partial P}\right)_T
	       =-{1 \over {P A }}{\biggl ({{\partial {\rm Re}(E)} \over
			{\partial P}}\biggr)_T}
                = c_0 + c_1 P + c_2 P^2 \cdots + c_n P^n + \cdots
\label{eq:cs}
\end{equation}
where the elastic material has cross-sectional area $A$. $c_0$ is the 
inverse of the normal bulk modulus, and the higher order terms $c_n$ 
represent nonlinear elastic coefficients. These are normally ignored, because by
the time they become important plastic deformation or fracture has set in, 
but in principle a careful experiment could measure them. Now, if $E(P)$
is an analytic function, then it obeys Cauchy's theorem,
\begin{equation}
E(-P)={1 /{2 \pi i}}\oint\limits_{\gamma}{{E(T)/{T+P}}} dT 
={1 /{2 \pi i}}\oint\limits_{\gamma}
	{{E(T)/T \sum_{n=0}^\infty{(P/T)^n}}} dT .
\end{equation}
with $\gamma$ circling $-P$ in the complex plane. Because it
divides by $i$, it can be used to write the real part of the
free energy $E(P)$ in terms of the imaginary part. Exchanging the sum
and the integral, choosing contours and checking asymptotics,
we can calculate the $n$-th nonlinear elastic constant in equation \ref{eq:cs}
asymptotically for large $n$:
\begin{equation}
c_n \rightarrow {(-1)^n (n+2) / \pi A} \int\limits_0^B{{{\rm Im} E(T)} / 
{T^{n+1}}} dT
\label{eq:cN}
\end{equation}
So because the (complex) energy is an analytic function, the high-order
nonlinear elastic constants are related to the fracture rate.

We used this to answer an extremely formal question. What is the radius of
convergence of nonlinear elastic theory? We can use the ratio
test: the radius of convergence equals the ratio ${c_{n} / c_{n+1}}$ 
as $n \rightarrow\infty$. Using
(\ref{eq:cN}) and (\ref{eq:ImE}) and doing the integrals, we find
\begin{equation}
{c_{n} / c_{n+1}}\rightarrow
        - n^{-1/2} {
\biggl({{{8 \beta Y \alpha^2}/{\pi (1-\sigma^2)}}\biggr)}^{1/2}
} \mbox{\hspace{0.1in}  as n} \rightarrow\infty.
\end{equation}
(Doing the prefactor carefully\cite{BuchelPRE} with various continuum limits
leaves this expression unchanged, except that the surface tension $\alpha$
becomes the temperature-dependent surface free energy.)
The ratio goes as $n^{-1/2}$, and the radius of convergence is zero.

Should we be concerned? Hooke's law is the first term of a series that doesn't
converge? First, the physical origin of this is that our brittle material
is in principle unstable as soon as the pressure $P$ becomes negative. If you're
not concerned about the fact that your bridges in theory will eventually 
fall down because of thermal crack nucleation, perhaps the associated 
non-convergence of Hooke's law shouldn't distress you. Indeed, thermal
fluctuations are probably too weak to make observable effects in any
measurement of the nonlinear elastic constants. Second, nonlinear elastic theory
has lots of company: quantum electrodynamics\cite{Dyson} (the most
quantitatively successful theory in physics) and Stirling's series
for $n!$ also have zero radius of convergence.

Intellectual curiosity aside, what of practical importance can we glean
from our calculation? First, perhaps the high-order nonlinear
elastic constants of disordered materials might be computed by
similar means, and (disorder being important
to the fracture rate) might have effects large enough to measure. Second, the
imaginary part of the frequency dependent elastic constants are related
to viscosities: perhaps considering both external strain and frequency
together might be illuminating (vibration-induced thermal fracture rates?)
Finally, the corresponding calculation for thermal nucleation of dislocation
loops would be straightforward, and might relate to the plastic flow rate at
very early stages (before work hardening sets in). The corresponding problem in
two dimensions has been well studied,\cite{2DDislocations} but
it seems likely that tangling of dislocations dominates in three dimensions.

\vskip 24pt

{\bf CRACK GROWTH LAWS}

\vskip 12pt

What are the crack growth laws? On long length and time scales, most 
macroscopic systems obey simple laws of motion, which depend on the properties
of the material only through setting constants (like density and elastic
constants) or functions (like surface energies including the crystalline
anisotropy). Can we deduce the laws describing the growth of a crack front,
coarse-graining over all the atoms, grains, precipitates, and plastic
deformation near the crack tip, to get relatively simple laws depending on
only a few materials parameters or functions?

Jennifer Hodgdon and I studied this problem, inspired by conversations with 
Tony Ingraffea and members of his group. Ingraffea's group\cite{IngraffeaWeb}
solves the hard part of the problem: given a complicated, three-dimensional
crack, they can tell us the stress intensity along the crack tip. All they
wanted was the rules for advancing the crack forward in time. The simplest
problem for us to study, and the one of most current interest to their
group, was quasistatic fracture in an isotropic, homogeneous medium. Including
crystalline anisotropy would lead us to the physics of cleavage along the
crystalline planes. Including random inhomogeneities into theories of
brittle fracture leads to logarithmically rough fracture 
surfaces\cite{RamanathanFisher1}, but not as rugged as the 
experiments\cite{Bouchaud}. Including the inertial dynamics of 
the growing front leads to interesting traveling waves along
the crack front\cite{FrontWaves} that might improve the agreement with 
experiment. In ductile and intergranular fracture, the roughness reflects
void growth and the presence of grains and inclusions.\cite{Anderson}
Our work did not include any of these effects.

Naturally, Ingraffea's group already had a working solution. For mode I
fracture, the material is described by a crack growth velocity $v(K_I)$.
For mode II fracture, the crack turns abruptly until it is mode I (the 
principle of local symmetry\cite{GS}). There were two other rules in
the literature for picking the crack growth direction: one maximizing the 
energy release\cite{Wu}, and one moving into the direction of minimum strain
energy density\cite{Sih}. All three rules seemed to predict crack paths
which agreed well with one another\cite{Wawrzynek}. At that time (around 1990)
they were just moving into 3D fracture, and were quite interested in more
crack growth laws for more general geometries.

Jennifer Hodgdon and I decided to try to write the most general crack growth
law allowed by symmetry. Let us describe the local crack front with an
orthonormal triad of vectors: $\bf \hat b$ perpendicular to the local crack
face, $\bf \hat t$ tangent to the crack front, and $\bf \hat n$
pointing in the direction along which the crack was last growing. Our
theory is valid on long length and time scales: in particular, we assume
that the curvatures of the the radii of curvatures for the crack surface
and crack front are large, so to zeroth order the crack is straight and
flat (we'll expand in the curvatures). There are
then three symmetries of the crack: the reflection $R_{\rm plane}$
in the $\bf \hat n - \hat t$ plane of the crack, the reflection $R_{\rm cross}$
in the $\bf \hat n - \hat b$ plane perpendicular to the crack front
$\bf \hat t$, and the 180$^\circ$ rotation about $\bf \hat n$ (which is
the product of the two other symmetries). 

The first convenient thing to fall out of the symmetry analysis is the 
decomposition of the strain field into modes. The linear elastic solution
near the crack tip can be decomposed into a superposition of solutions
with definite symmetries under $R_{\rm plane}$ and $R_{\rm cross}$. Mode I
is symmetric under both transformations, mode II is antisymmetric under 
reflections in the plane but symmetric to reflections perpendicular to the
front, and mode III is antisymmetric under both. Jennifer found that there
was yet another mode (mode IV?) which is symmetric under $R_{\rm plane}$ and
antisymmetric under $R_{\rm cross}$. Presumably Hodgdon's mode IV 
is usually ignored because it does not introduce a crack opening.

We then assume that the body being cracked and the radii of curvature
of the local crack front and crack surface are all large compared to the
inhomogeneities in the material and the nonlinear zone around the crack tip.
This is what is needed to ensure that there is a region around the crack
tip where the elastic displacement fields scale with the square root of the
distance $r$ to the crack tip. (In general\cite{Hodgdon,Ruina} the displacement
field for each mode can be expanded in a series of all half-integer powers
of $r$. The less singular terms are important only far from the crack, and
are determined by the boundary conditions and the crack shape; the more
singular terms are determined by the nonlinear zone around the crack tip, and
are also cut off by that zone.) Thus, for example, we describe
ductile fracture only if the body is much larger than the plastic zone.
We write the growth laws of the crack
in terms of the three stress-intensity factors $K_I$, $K_{II}$, and $K_{III}$
in this (so-called $K$-dominant) region, their gradients along the crack 
front, and the curvatures of the crack front and crack surface. We impose
the symmetries of the problem described above, plus a gauge 
symmetry\cite{SteveLanger,Hodgdon} associated with reparameterization of the
crack front, to find the most general law of motion allowed by symmetry.

In two dimensions, we found that the most general crack growth law was of the
form
\begin{equation}
\partial {\bf x}/\partial t = v(K_I,K_{II}^2,K_{III}^2) {\bf \hat n}, 
\hskip 1truein \partial {\bf \hat n}/\partial t = -f(K_I,K_{II}^2,K_{III}^2) 
			K_{II} {\bf \hat b},
\end{equation}
where $\bf x$ is the location of the crack tip,  $v$ is the velocity of the
crack, and $f$ represents the tendency for a crack to turn under the external
load (hence changing the direction $\bf \hat n$ pointing along the crack). 

What does this equation imply about the crack growth? If $K_{II}=0$, the crack
won't turn; otherwise it will turn (if $f>0$) so as to reduce $K_{II}$ in
magnitude. Thus we agree with the principle of local symmetry, except that
our cracks will turn gradually towards this direction, rather than jump
abruptly to the new orientation. Over what turning radius does our crack
turn? Using results of Cotterell and Rice\cite{Cotterell}, Hodgdon showed
that if the angle of the crack differs from the angle that makes $K_{II}=0$
by a small amount $\Delta\theta$, then $K_{II}=K_{I}\Delta\theta/2$;
this implies that $\Delta\theta ~ \exp(-f K_I x/2v)$, and the angle decays
exponentially to pure mode I with a material-dependent decay length of
$2 v / f K_I$. We expect this length scale to be set by a microscopic scale
characteristic of the material: the atomic size in a glass, the size of the
nonlinear zone in a ductile material, etc. This means, in essence, that our
analysis agrees with the traditional prescription that the crack turns
abruptly. We think of writing the crack growth laws as differential equations
replaces the atomistic short-distance cutoff with a smooth one.

Our analysis did not make any assumptions about microscopic mechanisms.
We believe that arguing whether the crack turns to maximize the energy 
release or minimize the strain energy density is misguided: if you turn the 
crack according to the wrong rule, in the succeeding step it'll keep turning
and you'll converge to the same final trajectory (making $K_{II}=0$) with
a radius of curvature given by the step size in your algorithm.

So much for reinventing what's already known. What about three dimensions?
Now everything depends on where we are along the crack front. If we let $s$
measure arclength along the crack front, then up to first order in gradients
\def\ddx#1#2{{\partial #2\over\partial #1}}
\def\kI{K_{\rm I}}
\def\kII{K_{\rm II}}
\def\kIII{K_{\rm III}}
\def\nhat{\hat n}
\def\bhat{\hat b}
\def\that{\hat t}
\def\xvec{\vec x}
\def\dadotb#1#2{ {\partial #1 \over \partial s} \cdot #2 }
\begin{eqnarray} 
 \partial x / \partial t =& v \nhat + w \that \cr	
 \partial \nhat / \partial t =& - \left[ \ddx{s}{v} +
   w \ddx{s}{\that}  \cdot \nhat \right] \that + 
  \left[ - f \kII + g_{\rm I} \kIII \ddx{s}{\kI} +
   g_{\rm II}\kII \kIII \ddx{s}{\kII} + g_{\rm III}\ddx{s}{\kIII} + \right. \cr
&\qquad \left.  h_{tb} \dadotb{\that}{\bhat} +
   h_{nt} \kII \dadotb{\nhat}{\that} +
 (h_{nb}\kII\kIII + w)  \dadotb{\nhat}{\bhat} \right] \bhat ,\cr 
\label{eq:3D}
\end{eqnarray}
where $f$, $g_{\alpha}$, and $h_{ij}$ are functions of $\kI$, $\kII^2$, and
$\kIII^2$, and the velocity $v$ can be a function of these and a number of
gradient terms\cite{Hodgdon}. Only $f$ among these constants, which gives
the turning radius for mode $II$ fracture, is expected to be large (as
discussed above): the others should be all of the same order of magnitude.

What can we do with this expression? 
First, if the crack begins nearly flat and straight, we can do perturbation
theory to see if the crack becomes flatter and straighter, or if it goes
unstable.
Jennifer Hodgdon\cite{HodgdonThesis} found that cracks under mode I are stable
(as expected), but that cracks under mixed mode I and mode III (twisting)
can be stable or unstable depending on the materials constants: in particular,
if $g_I>0$, steady-state mode III cracks are unstable to small perturbations.
Experimentally, she found that cracks under mode III turned to form a helical
crack surface: such turning wasn't allowed for in our perturbation theory, which
assumed periodic boundary conditions. Others find, for longer cracks, that
mode III fracture is unstable to the formation of a ``factory roof''
morphology. It wasn't possible at the time to test Hodgdon's results with
realistic crack-growth simulations, and this work remains 
unpublished.\cite{HodgdonThesis}

Second, we can use it to propagate real 3D cracks in simulations, and compare
to experiments and more microscopic simulations. I anticipate that our 
symmetry analysis will prove important and useful, but that there will be
additional internal variables (perhaps crack-tip curvature\cite{Argonne}) that
will be necessary for describing the crack, and which may induce crack branching
and other effects missing in our existing theory. 


\end{document}